\documentclass[12pt]{article}

\usepackage[]{graphicx,bm,color,subfigure,amsmath}

\usepackage{scicite}

\usepackage{times}
\usepackage{textcomp}
\usepackage{gensymb}
\usepackage{xr}
\usepackage{placeins}

\newenvironment{sciabstract}{%
\begin{quote} \bf}
{\end{quote}}

\newcommand{\beginsupplement}{%
        \setcounter{table}{0}
        \renewcommand{\thetable}{S\arabic{table}}%
        \setcounter{figure}{0}
        \renewcommand{\thefigure}{S\arabic{figure}}%
     }

\newcounter{lastnote}

\title{Two-dimensional mutual synchronization in spin Hall nano-oscillator arrays}

\author
{M. Zahedinejad$^1$, A. A. Awad$^{1,2}$, S. Muralidhar$^1$, R. Khymyn$^1$,  \\ H. Fulara$^1$, H. Mazraati$^{2,3}$, M. Dvornik$^{1,2}$, and J. \AA kerman$^{1,2\ast}$\\ 
\\
\normalsize{$^{1}$Physics Department, University of Gothenburg, 412 96 Gothenburg, Sweden}\\
\normalsize{$^{2}$NanOsc AB, Kista 164 40, Sweden}\\
\normalsize{$^{3}$Department of Applied Physics, School of Engineering Sciences,}\\ 
\normalsize{KTH Royal Institute of Technology, Electrum 229, 164 40 Kista,Sweden}\\
\\
}
\date{}

\begin{document} 


\baselineskip24pt


\maketitle


\begin{sciabstract}
 
Spin Hall nano-oscillators (SHNOs) utilize pure spin currents to drive local regions of magnetic films and nanostructures into auto-oscillating precession. If such regions are placed in close proximity to each other they can interact and sometimes mutually synchronize, in pairs or in short linear chains. Here we demonstrate robust mutual synchronization of two-dimensional SHNO arrays ranging from 2 x 2 to 8 x 8 nano-constrictions, observed both electrically and using micro-Brillouin Light Scattering microscopy. The signal quality factor, $Q=f/\Delta f$, increases linearly with number of mutually synchronized nano-constrictions ($N$), reaching 170,000 in the largest arrays. While the microwave peak power first increases as $N^2$, it eventually levels off, indicating a non-zero relative phase shift between nano-constrictions. Our demonstration will enable the use of SHNO arrays in two-dimensional oscillator networks for high-quality microwave signal generation and neuromorphic computing.

\end{sciabstract}



The interest in bio-inspired oscillatory computing \cite{borisyuk2000biosys,pikovsky2015chaos,jaeger2004sci,grollier2016ieee,pufall2015ieee,locatelli2014ntmat} 
is rapidly increasing in an effort to mitigate the inevitable end of Moore's law \cite{xu2018ntelec}. Although neuronal activities may seem slow, our brain is the most efficient processing device for cognitive tasks with regards to energy consumption and speed thanks to its massively interconnected oscillatory neurons \cite{buzsaki2006rhythms}. In many biological neuronal networks, such as the pacemaker cells in the heart \cite{glass2001nature} and the hippocampus and the cortex \cite{gregoriou2009sci,veit2017ntneusci,penn2016pnas,buzsaki2004sci} synchronization of oscillatory neurons plays a central role. While recent memristive \cite{kumar2017nt,ignatov2017sci,pickett2013ntmat},
superconducting \cite{segall2017pre,galin2018pra}, optical
\cite{inagaki2016science,mcmahon2016science}, and micromechanical \cite{fang2016SciAdv,shim2007sci} oscillator arrays have been demonstrated, realizing a large physical network that meets technical requirements such as room temperature operation, scaling, integration, high speed, and low power consumption remains a challenge.

Spin transfer torque nano-oscillators (STNOs) are one of the most promising candidates addressing these requirements \cite{csaba2013ieeemag,yogendra2016ieee}. Free running STNOs can interact with other STNOs to synchronize via coupling mechanisms that can be electrical\cite{vodenicarevic2017scirep,tsunegi2018scirep}, exchange based\cite{ruotolo2009ntn}, dipolar \cite{Locatelli2015}, or due to spinwave propagation\cite{madami2011ntn, kaka2005nt,mancoff2005nt,Pufall2006prl,sani2013ntc, Houshang2015ntn}, and this way deliver higher power and more coherent microwave signals. A recent study has highlighted a comparable performance of one STNO\cite{torrejon2017nt} in vowel recognition rate compared to state-of-the-art CMOS. Later, a learning process was demonstrated by a linear chain of four electrically synchronized STNOs implementing reservoir computing \cite{romera2018nt}. However, in order to allow such networks to process more complicated tasks, many more oscillators must interact and mutually synchronize. A substantial increase in the number of STNOs within the demonstrated chains is not feasible as their synchronization bandwidth is limited and their device variability large \cite{romera2018nt}. 

Spin Hall nano-osillators (SHNOs) have recently emerged as an attractive alternative to STNOs \cite{demidov2012ntm,Ranjbar2014ieeeml, Chen2016} as they are directly compatible with CMOS \cite{zahedi2018apl} and can be mutually synchronized in easy to fabricate chains \cite{awad2017ntphys}. Here we demonstrate that nano-constriction SHNOs can also be mutually synchronized in two-dimensional arrays comprised of as many as 64 SHNOs. 
Each oscillator (neuron) within the 2D array interacts with its neighbors via both exchange and dipolar coupling, which can furthermore be tuned by both the drive current, and the strength and direction of the magnetic field. As expected from theory \cite{Kim2008prl}, the signal quality factor of the mutually synchronized state increases linearly with number of SHNOs, reaching $Q$ = 170,000 for 64 SHNOs, \emph{i.e.}~an order of magnitude higher than the previous best literature values.




\begin{figure}
\begin{center}
\includegraphics[width=1\textwidth]{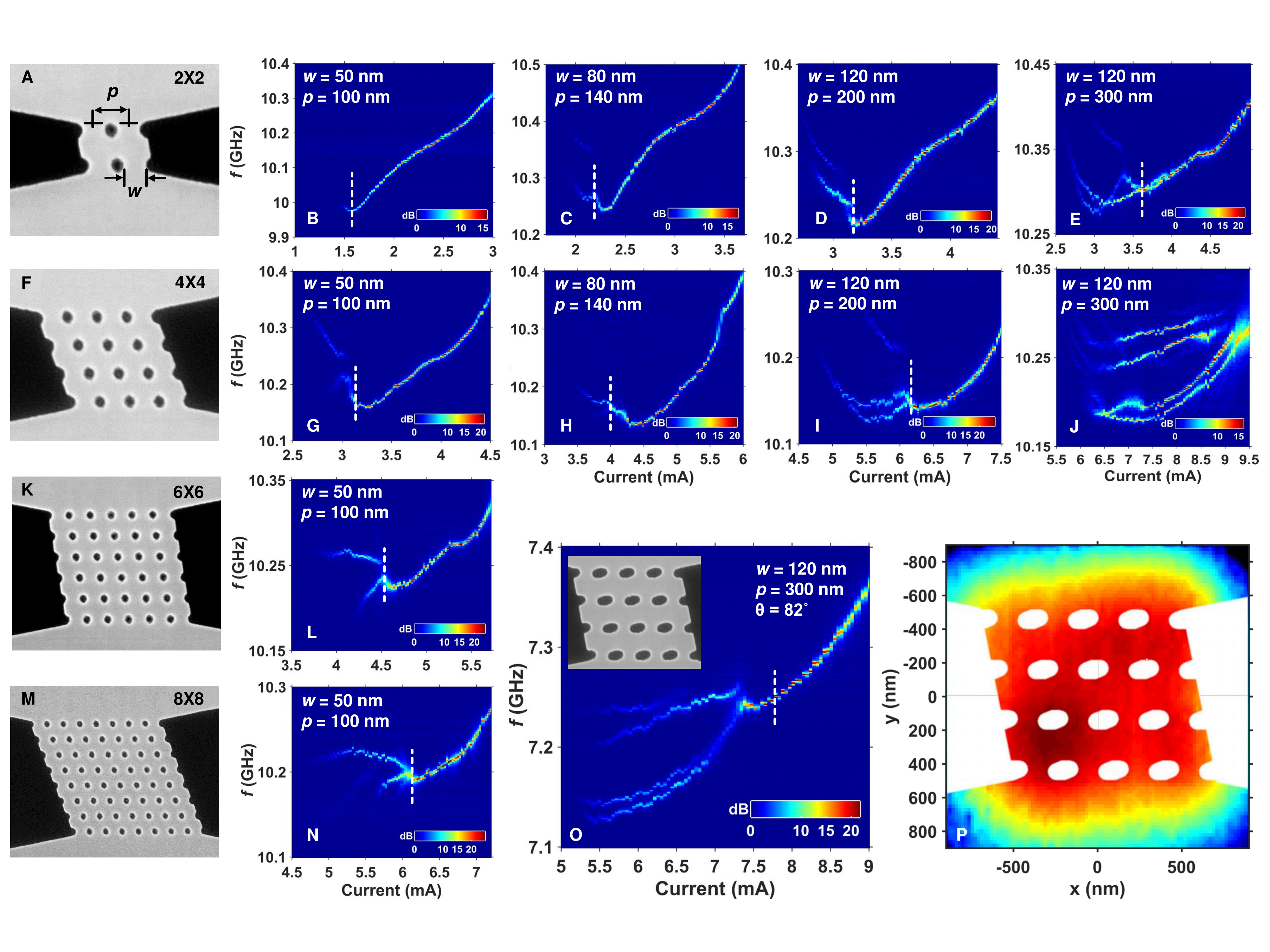}
\caption{\textbf{Power spectral density (PSD) and micro-Brillouin Light Scattering microscopy of spin Hall nano-oscillator arrays.} All data was acquired in a magnetic field of 0.68 T with an out-of-plane angle of 76$^{\circ}$ and an in-plane angle of 30$^{\circ}$. (\textbf{A}) Scanning electron microscope image of a 2x2 SHNO array showing the definition of the width ($w$) and the pitch ($p$). (\textbf{B} to \textbf{E}) PSD of four different 2x2 SHNO arrays with different constriction widths (50, 80, and 120 nm) and SHNOs center to center distance as pitch size (100, 140, 200, 300 nm). White dashed lines indicate the current where a robust mutually synchronized state is achieved. (\textbf{F}) SEM picture of a 4x4 SHNO array. (\textbf{G} to \textbf{J}) PSD of four different 4x4 arrays having the same width and pitch as the 2x2 arrays in \textbf{B} to \textbf{E}. Mutual synchronization is only observed in the first three SHNO arrays. (\textbf{K}) and (\textbf{M}) SEM pictures of a 6x6 and an 8x8 SHNO array. (\textbf{L}) and (\textbf{N}) PSD of the smallest 6x6 and the smallest 8x8 SHNO array, both showing robust mutual synchronization. (\textbf{O}) PSD of the same 4x4 array as in \textbf{J} with the out-of-plane field angle increased from 76$^{\circ}$ to 82$^{\circ}$ to increase the coupling between SHNOs. The inset shows the SEM image of the defined array. (\textbf{P}) Micro-Brillouin Light Scattering microscopy image of the 4x4 array in \textbf{O} obtained at $I=$ 7.6 mA (white dashed line) operating point confirming the complete mutually synchronized state.}
\label{fig:1}
\end{center}
\end{figure}

Fig.\ref{fig:1}A shows an SEM image of a 2 x 2 SHNO array with the design parameters $w$ for width, and $p$ for pitch (\emph{i.e.}~nano-constriction separation) defined. The array has a slight tilt angle to better accommodate for the 30$^{\circ}$ in-plane angle of the applied field, since the auto-oscillating regions extend outwards in a direction perpendicular to this angle. We fabricated 24 different square SHNO arrays where the number of nano-constrictions was varied from 2x2 to 10x10 (\emph{i.e.}~from 4 to 100) and the following four pairs of $w$ and $p$ were chosen: $(w$,$p)$ = $(50,100)$, $(80,140)$, $(120,200)$, and $(120,300)$; all numbers in nanometers. Fig.\ref{fig:1}F, Fig.\ref{fig:1}K, and Fig.\ref{fig:1}M in the same column show the corresponding SEM images for a 4x4, 6x6, and 8x8 array of  $(w$,$p)$= $(120,200)$ respectively. For details of the fabrication process, see supplementary material \ref{num1}.

Fig.\ref{fig:1}B-E show the power spectral density (PSD) for the four different 2x2 arrays as a function of total drive current through the array. All PSDs show the typical non-monotonic current dependent frequency as the nano-constriction edge mode expands with current \cite{Awad2017, dvornik2018pra}. All four arrays exhibit mutual synchronization at a synchronization current which increases linearly with $w$. (Measurement details in supplementary material \ref{num2})

Fig.\ref{fig:1}G-J show the corresponding PSD vs.~current for the four different 4x4 arrays. While the overall non-monotonic current dependence of the frequency is approximately the same as in the 2x2 arrays, mutual synchronization is now only achieved in the three first arrays. The 4x4 arrays with 300 nm separation between the nano-constrictions instead show four distinct individual signals, all with approximately the same linewidth and peak power. The four distinct signals indicate that the array exhibits partial mutual synchronization, \emph{i.e.}~the array synchronizes along its four chains, but the four chains do not synchronize with each other. The coupling strength is hence different along the chains and in between chains. The behavior of the 5x5 array is essentially identical with the 4x4 array, with complete synchronization in the first two arrays, but only chain wise synchronization at 200 nm and 300 nm separation (see Supplementary material \ref{sup2}).

Fig.\ref{fig:1}L\&N show the PSD of the 6x6 and 8x8 arrays at the smallest dimensions. At all larger dimensions, neither the 6x6 nor the 8x8 array showed complete mutual synchronization (see Supplementary material \ref{sup2}). Similarly, the 10x10 arrays did not show complete mutual synchronization at any dimension.

As discussed in Ref.~\cite{dvornik2018pra}, the auto-oscillating regions can be expanded by increasing the out-of-plane angle of the applied field. This should allow us to increase the coupling strength between neighboring constrictions and hence potentially allow us to synchronize larger arrays. In Fig.\ref{fig:1}O we show a 4x4 array similar to the one in Fig.\ref{fig:1}J but measured at an increased field angle of 82$^{\circ}$. The four signals from the individually synchronized chains now merge into a single signal at about 7.3 mA, confirming the important role of the out-of-plane field angle. At a separation of 300 nm it then becomes meaningful to explore the spatial profile of the synchronized state using Brillouin Light Scattering microscopy, as the separation is large enough for any variation within the array to be resolved. Fig.\ref{fig:1}P shows that the entire 4x4 array is energized with a relatively uniform spin wave intensity throughout the array.  
\begin{figure}
\begin{center}
\includegraphics[width=5in]{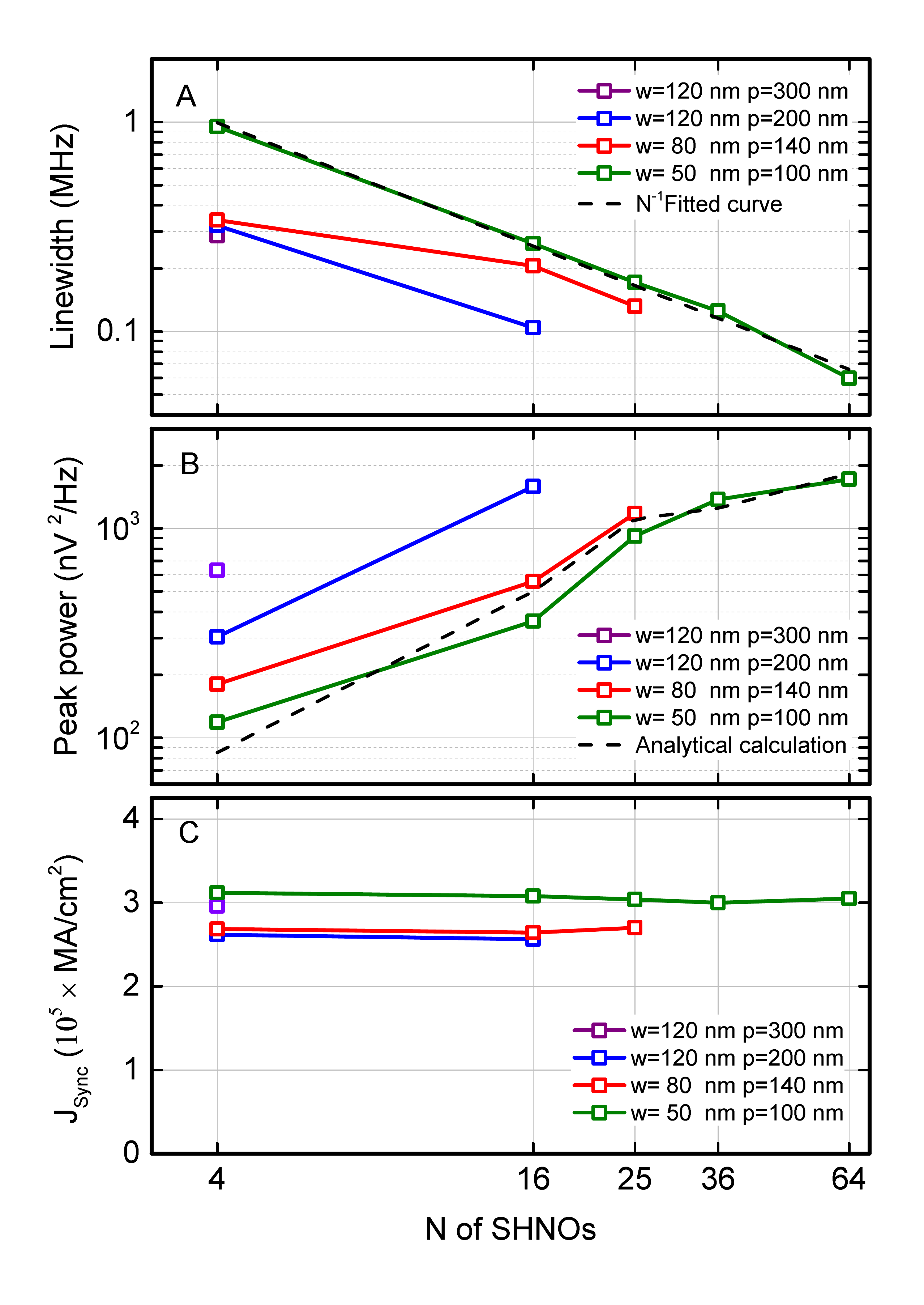}
\caption{\textbf{Linewidth, peak power and synchronization current density analysis of SHNO arrays.} (\textbf{A}) Linewidth of arrays of different $w$ and $p$ plotted only for those which reach robust synchronization. Black dashed line shows predicted linewidth scaling of proportional to N$^{-1}$. (\textbf{B}) Peak power values measured for all synchronized points shown in \text{A}. Black dashed line indicates the analytical calculation of peak power for arrays with $(w$,$p)$= $(50,100)$ considering the phase difference. (\textbf{C}) Synchronization current density for arrays with different $(w$,$p)$.}
\label{fig:2}
\end{center}
\end{figure}

In Fig.\ref{fig:2} we summarize the microwave signal properties of all arrays showing complete mutual synchronization. The linewidth plotted in Fig.\ref{fig:2}A was chosen as the lowest value observed at a certain current in the mutually synchronized regions for ten consecutive measurements. We found this approach to yield consistent values and a consistent trend between arrays since the frequency showed a tendency to slowly wander over a much wider range than its intrinsic linewidth. When comparing the individual spectrum analyzer measurements most of them would yield the same narrow linewidth (the lowest value), albeit at different central frequencies, but sometimes the measurement would be artificially broad due to the slow movement of the central frequency. The origin of this extrinsic second source of linewidth is yet unclear.

The black dashed line is a fit to $N^{-1}$, which clearly shows how the intrinsic linewidth decreases in inverse proportion to the number of mutually synchronized constrictions. This is consistent with the total mode volume, or total energy of the auto-oscillation state, increasing linearly with $N$.\cite{Kim2008prl} This also explains why the linewidth improves with increasing nano-constriction width.

Fig.\ref{fig:2}B shows the highest peak power at any current in the mutually synchronized regions. As the total (integrated) microwave power in the mutually synchronized state should increase linearly with $N$ for a square array \cite{georges2008apl} we expect the peak power to increase as $N^2$. While this is indeed observed for small $N$ = 4--25 for all arrays, the peak power eventually levels off for larger $N$ = 36--64. This roll off behavior can be explained by introducing a small relative phase difference, either between individual constrictions or between individual chains. While this phase shift is negligible in small arrays, it will add up to a significant phase shift between the constrictions or chains farthest apart in the larger arrays. As they will then no longer add their voltages purely in phase, the $N^2$ scaling will no longer hold; the net power can eventually decrease all the way back to zero as the array size grows even larger. It is hence crucial to keep any relative phase shift to a minimum to also maximize the output power.

The dashed black line in Fig.\ref{fig:2}B is a calculation of the expected peak power using the measured resistance values for each array and the assumption that there is a chain-to-chain relative phase shift of 16$^{\circ}$. The calculation agrees quite well with the experimental dependence on $N$ and indicates that the peak power would not increase substantially for 10x10 and larger arrays (see supplementary material \ref{sup3} for details).

We finally show how the synchronization current density ($J_{Sync}$) depends on SHNO array dimensions (Fig.\ref{fig:2}C). We first note that $J_{Sync}$ is virtually independent on number of SHNOs in the array as long as $w$ and $p$ stay the same. For the two arrays with $w$=120 nm, we can compare the impact of pitch and conclude that larger separation requires a slightly higher current density for mutual synchronization. However, the overall trend is that $J_{Sync}$ is rather independent on array dimensions at this level of detail.

The remarkably low linewidh of $\sim$60 kHz at an operating frequency of $\sim$10 GHz, leads to quality factors as high as $Q=f/\Delta f$ = 170,000. It is noteworthy that this is the first time that $Q$ of any spin based nano-oscillator has surpassed the previous record of 18,000 reported in 2004 \cite{rippard2004prb}. As the $N^{-1}$ dependence of the linewidth does not show any sign of levelling off for higher $N$, mutual synchronization of yet larger arrays can be pursued to further improve $Q$. The very low linewidth will greatly simplify the design of phase-locked loops (PLL) to further stabilize the microwave signal \cite{tamaru2015scirep}.

To truly benefit from the improved microwave signal coherence, it will however be important to increase the output power by other means than synchronization alone. As the best output power to date has been reported in mutually synchronized magnetic tunnel junction (MTJ) based STNOs \cite{tsunegi2018scirep}, it will be necessary to use MTJs also in SHNOs, ideally fabricated directly on top of the auto-oscillating region. While all SHNO arrays used Pt/NiFe bilayers in our study, MTJ based SHNOs should be possible to fabricate using \emph{e.g.}~W/CoFeB based bilayers \cite{zahedi2018apl}, where a W/CoFeB/MgO/CoFeB stack could be used to fabricate an MTJ on top of the constriction region. The resulting separation of the metal based SHNO drive and mutual synchronization on the one hand, from the MTJ based SHNO sense path on the other, will allow the use of much higher resistance-area product MTJ stacks with much higher tunneling magnetoresistance \cite{liu2012prl}.

In order to perform classification or segmentation (e.g. image processing, and vowel recognition) of big data, it is crucial to scale up the network as the maximum number of distinguishable classes or segments depend on the number of sufficiently interacting oscillators to provide a useful mix of partially or fully synchronized states. Our arrays offer a proper scaling path, currently accommodating 64 fully synchronized SHNOs and 100 partially synchronized SHNOs, all operating at tens of gigahertz frequencies requiring only nanoseconds to reach stable synchronized states. Their footprint is tiny (750 nm x 750 nm at present), and there are no practical issues to shrink the network size even further \cite{durrenfeld2017nanoscale}. Multi-core networks hosting many SHNO arrays could potentially handle much larger input data sets in parallel to speed up the processing even further.

In conclusion, we have fabricated the first two-dimensional SHNO arrays with up to $N$ = 100 nano-constrictions and demonstrated robust mutual synchronization in arrays with up to $N$ = 64 constrictions. We find that the linewidth scales inversely with the number of mutually synchronized constrictions and can reach below 60 kHz at frequencies of about 10 GHz. The resulting quality factors are an order of magnitude higher than the the best literature values. Our demonstration will enable the use of SHNO arrays in two-dimensional nano-oscillator networks for high-quality microwave signal generation and neuromorphic computing.

\bibliography{scibib}

\bibliographystyle{Science}

\section*{Acknowledgments:} This work was supported by the European Research Council (ERC) under the European Community’s Seventh Framework Programme (FP/2007-2013)/ERC Grant 307144 “MUSTANG.” This work was also supported by the Swedish Research Council (VR), the Swedish Foundation for Strategic Research (SSF), and the Knut and Alice Wallenberg Foundation.
\newline

\clearpage


\beginsupplement

\section*{Supplementary materials}

\section{Materials and Methods}\label{sup1}
\subsection{Sample fabrication}\label{num1}
To fabricate SHNO arrays, a bi-layer of Ni$_{80}$Fe$_{20}$(3nm)/Pt(5nm) was deposited at room temperature on an 18~mm$\times$18~mm high resistivity silicon substrate (10 k$\Omega$.cm). The native oxide layer on the substrate was removed by plasma cleaning before the deposition process. A magnetron sputtering machine with the base pressure lower than 3$\times$10$^{-8}$~Torr was used while the Ar pressure was kept at 3 mTorr during the deposition. The sample was covered by 37nm of Hydrogen silsesquioxane (HSQ) electron beam resist and SHNO arrays were written in HSQ by a Raith EBPG 5200 electron beam lithography machine operating at 100 kV. The patterns were then transferred to the bilayer by ion beam milling process at $30^\circ$ ion incident angle with respect to the film normal to minimize sidewall redepositions. To define the top coplanar waveguide (CPW) contact for dc and microwave measurements, optical lithography was performed followed by HSQ removal only at contact areas in diluted buffered Hydrofluoric acid. Finally, a 1 $\mu$m thick layer of Cu(980nm)/Pt(20nm) was deposited, and CPWs were obtained after resist removal by the lift-off process.

\subsection{Microwave characterization}\label{num2}
We used a custom-built probe station where the stage can rotate the sample holder between the poles of an electromagnet to apply an out-of-plane magnetic field to the sample. The in-plane and out-of-plane angles of the sample were fixed at 30$^\circ$ and 76$^\circ$, respectively while the magnetic field was set to $\mu_0 H$= 0.68 $T$ for all measurements. A  direct current was applied through the dc port of a bias-T to excited auto oscillation in  SHNO array while emitted microwave signal from array was picked up by the high frequency port of bias-T and was sent to a low noise amplifier before it was recorded by a high frequency spectrum analyzer (SA). The recorded spectra were then corrected to correspond to the power emitted by the device, taking into account the amplifier gain, the losses from the rf components and cables, and the impedance mismatch between the device and the 50 $\Omega$ measurement line and load. The auto-oscillation linewidth and peakpower were extracted by fitting a single symmetric Lorentzian function.

\subsection{$\mu$-BLS characterization}\label{num3}
The magneto-optical measurements were performed using room temperature micro-focused BLS measurements. Spatially resolved maps of the magnetization dynamics are obtained by focusing a polarized monochromatic 532 nm single frequency laser (solid state diode-pumped) using a high numerical aperture (NA=0.75) dark-field objective, which yields a diffraction limited resolution of 360 nm. The scattered light from the sample surface is then analysed by a high-contrast six-pass Tandem Fabry-Perot interferometer TFP-1 (JRS Scientific Instruments). The obtained BLS intensity is proportional to the square of the amplitude of the magnetization dynamics at the corresponding frequency. 

\section{Electrical measurements}\label{sup2}
\begin{figure}[h]
\begin{center}
\includegraphics[width=1\textwidth]{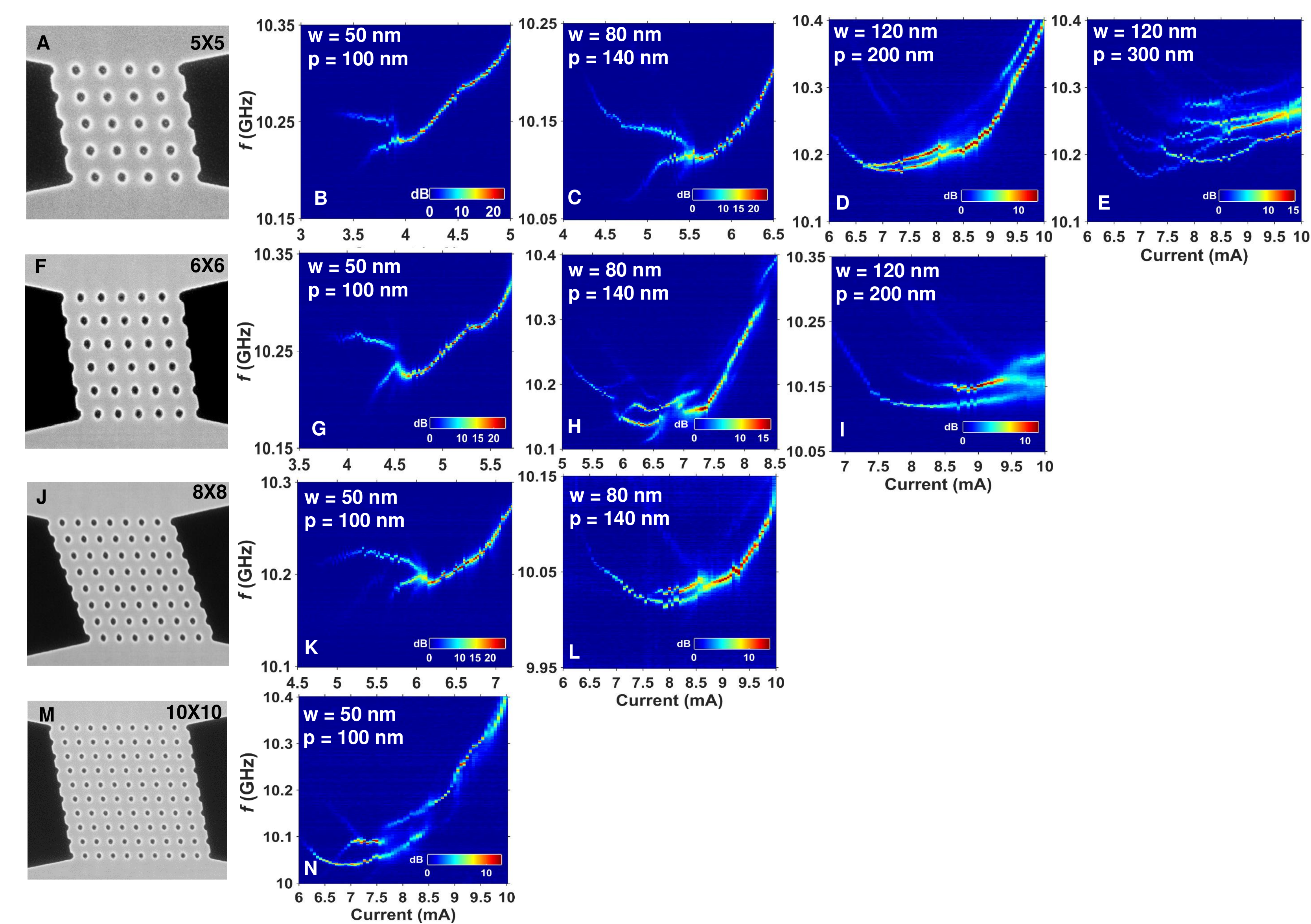}
\caption{\textbf{Electrical measurements for large arrays which partially synchronize, but fail to synchronize as an ensemble.}}
\label{fig:S1}
\end{center}
\end{figure}

The electrical measurements of larger arrays which lead to partially synchronized and/or non-synchronized states are shown in Fig.\ref{fig:S1}A to Fig.\ref{fig:S1}N.
\FloatBarrier

\section{Estimated power spectral density}\label{sup3}
We consider the power delivered by the array of $K \times M = N$ oscillators to the load $R_l$, taking into account the finite resistance of the mesa $R_m$. For the perfectly synchronized state, without any phase shift between oscillators, the power reads as \cite{georges2008apl}:
\begin{equation}
    P_N=\left[\frac{K M I_{dc} \Delta R_{ac}}{K R_c +M(R_l+R_m)}\right]^2 R_l,
\end{equation}
where $M$ defines the quantity of parallel branches with $K$ serial oscillators in each of them and, thus, for our case $K=M=2, 4, 5, 6, 8$. $R_l = 50 \Omega$ - load resistance, $R_m = 200 \Omega$ is the resistance of the sample outside the array region (mesa), $R_c = 80 \Omega$ - the resistance of each nano-constriction, $I_{dc}$ - dc current applied to each oscillator and $R_{ac}$ is the alternate resistance created by the magnetization precession through AMR, which we assume as a fitting parameter. 

Please note, that the above expression results in the proportionality of the delivered power to the total number of oscillators $N$ for the square arrays, when $K=M$. Since the measured power strongly deviates from such a scaling, we take into account the phase shift $\phi$ between neighboring columns. In this case, assuming the Lorentzian shape of the PSD with the linewidth $\Delta f$ (FWHM), one can write the maximum value of PSD:
\begin{equation}
    \text{PSD}_{max}=\frac{2}{\pi \Delta f}\left[\frac{K I_{dc} \Delta R_{ac}}{K R_c +M(R_l+R_m)}\sum_{j=0}^{M-1}\cos j \phi \right]^2  R_l^2,
\end{equation}
which is shown by a dashed line on Fig. \ref{fig:2} B with the fitted value $\phi=16.4 \degree$. Please note, that in the synchronized state the linewidth $\Delta f$ should be inversely proportional to the total power of auto-oscillations, i.e. to the total number $N$, and does not depend on the phase shift $\phi$ \cite{wiesenfeld1994apl}.

\clearpage

\end{document}